\documentclass[prb,preprint,aps,amsmath,amssymb]{revtex4}
\usepackage{graphicx}
\usepackage{dcolumn}
\usepackage{bm}
 \begin{document}
 \title{Response to ``Comment on `Performance of a spin-based insulated gate field effect transistor' [Appl. Phys. Lett. 88, 162503 (2006)]''}
\author  {Michael E. Flatt\'e}
\affiliation {Optical Science and Technology Center and Department of Physics and Astronomy, University of Iowa, Iowa City,
IA 52242, USA}\author{Kimberley C. Hall}
\affiliation{Department of Physics, Dalhousie University, Dalhousie, Canada}
\maketitle

In our previous Letter we analyzed the fundamental physical properties limiting the performance of spin field effect transistors and compared them to ordinary (charge-based) field effect transistors.\cite{APL} The comment by Bandyopadhyay and Cahay\cite{comment} makes two claims: (1) that the threshold voltage of a metal-oxide-semiconductor-field-effect-transistor (MOSFET) can be lower than the $400$~mV described in the semiconductor roadmap\cite{ITRS:2003}, and (2) that the injection of spin-polarized currents of sufficient polarization may not be possible. In this response we reiterate that the subthreshold voltage swing at room temperature precludes the achievement of (1) without sacrificing the on-off current ratio, and that no fundamental limits exist to the polarization of injected spin-polarized current.

In Ref.~\onlinecite{APL} our comparative analysis of the spin field effect transistor of Ref.~\onlinecite{HF} and the MOSFET was performed at room temperature, whereas in Ref.~\onlinecite{comment} the analysis was performed at zero temperature.  The textbook minimum subthreshold voltage swing of the MOSFET (see, e.g. Ref.~\onlinecite{MOSFET}) at room temperature is $60$~mV/decade. It is determined by thermal activation over the MOSFET barrier ($\exp[eV/k_BT]=10$, where $e$ is the electron charge, $V$ the voltage, $k_B$ Boltzmann's constant, and $T$ the temperature).
At room temperature the threshold voltage of 15.7~mV for the MOSFET from Ref.~\onlinecite{comment}  directly implies an on-off current ratio of 1.9, which does not compare favorably with the on-off ratios of $10^5$ for the spin-FET described in our Letter. Thus the statement of Ref.~\onlinecite{comment} that ``The MOSFET could always win by lowering the carrier concentration in the channel'' is false at room temperature. 

As described in detail in Ref.~\onlinecite{APL}, a major advantage of the spin-FET is the ability to avoid the fundamental MOSFET limit on threshold voltage due to the $60$~mV/decade subthreshold voltage swing. This is possible because the spin-FET switching is not performed by raising or lowering a barrier (as shown in Figs.~1~and~2 of Ref.~\onlinecite{APL}).

Ref.~\onlinecite{comment} inquires how much voltage drop is across the 2018 CMOS MOSFET oxide, which is directly available from the roadmap\cite{ITRS:2003}  (in 2018 it is about 70\%\ of the total).
The voltage drop across the oxide in Ref.~\onlinecite{APL} was neglected because the spin-FET's capacitance is very small ($\sim 9$ times smaller than the capacitance of the 2018 CMOS MOSFET oxide). Thus if the same oxide is used for the spin-FET as  planned for 2018 CMOS, the presence of the oxide changes the total spin-FET capacitance by only $\sim 10$\%, and the voltage drop across the oxide is $\sim 10$\% of the total voltage drop, changing the threshold voltage by $\sim 10$\%. 

The second criticism of Ref.~\onlinecite{comment} is that the injection of spin-polarized currents of sufficient polarization may not be possible. Ref.~\onlinecite{comment} declares that ``realistically, the on-off ratio will be low, probably $<10$ at room temperature'', relying on 2006 state of the art spin injection efficiencies. As stated in our Ref.~\onlinecite{APL}, our focus was on fundamental power dissipation limits, and the comparison presented was to 2018 CMOS. Ref.~\onlinecite{comment} has failed to provide any argument why the 2006 state of the art should be considered a fundamental limit on spin injection efficiency and apply to 2018 comparisons (indeed, 7 years ago spin injection into semiconductors had not even been demonstrated, yet now the efficiency at room temperature exceeds 70\%)\cite{Salis}. Just as advances in contacting technology of metals to semiconductors permitted nearly 100\%\ electron current injection (as opposed to a mixture of electron and hole current),  nearly 100\% spin-polarized currents are achievable through advances in ferromagnetic contacting technology. As addressed in Ref.~\onlinecite{APL}, the use of highly spin-selective barriers\cite{Salis} or half-metallic ferromagnetic contacts\cite{HM} are two avenues predicted to provide nearly 100\% spin injection efficiencies. No fundamental roadblocks prevent this.

Thus the criticism of Ref.~\onlinecite{comment} regarding the threshold voltage of the spin-FET is specious, as it compares a zero temperature result for the MOSFET with a room temperature result for the spin-FET. In our Letter the proper comparison was done. The criticism of Ref.~\onlinecite{comment} regarding the efficiency of spin injection is a guess, not based on physical analysis, that the rapid progress over the last 7 years towards high-efficiency spin-polarized current injection will suddenly stop. Fundamental analysis suggests otherwise.

This work was supported by DARPA/ONR and the Natural Sciences and Engineering Research Council of Canada.

\vfill\eject

\end{document}